# CFD SIMULATION OF X-ADS DOWNCOMER THERMAL STRATIFICATION

V. ANISSIMOV [(1)], A. ALEMBERTI [(2)]
[(1)] ENEA Bologna – Research Fellowship Holder
[(2)] Ansaldo Energia S.p.A – Divisione Nucleare

**Abstract**

This work presents a numerical simulation using CFX 4.4 of the Energy Amplifier Experimental Facility (X-ADS) downcomer channel. The simulation is focused on the Steady-State Analysis.

The Intermediate Heat exchangers (IHX) of the X-ADS reference configuration are immersed in the lead-bismuth eutectic of the downcomer. Due to the absence of a physical separation between the primary coolant hot and cold collectors, two different flow paths are available in the downcomer region: inside the IHX and outside it (IHX by-pass flow). The amount of IHX by-pass flow is determined by the balance between the driving force due to buoyancy (originated by the weight difference between the cooled fluid inside the IHX and the hot outside downcomer fluid) and the IHX pressure losses. At the IHX exit the two flow paths are mixed before the core inlet. This fact provides a potential for a downcomer thermal stratification, which is influenced by the actual value of coolant flow rate outside the IHX.

The amount and extension of the thermal stratification phenomena is the object of this study.

The simulation allowed studying the position and intensity of the thermal stratification phenomena. Several runs have been performed.

However, to limit the extension of the paper, we deal with the results of only one of the calculations performed, which resulted in the worst condition from the point of view of thermal loads on the structure. Other results obtained will be shortly recalled when needed.

## 1. THE X-ADS DOWNCOMER SUB-SYSTEM

The simplified scheme of the primary coolant flow paths is shown in Fig. 1.

Only one IHX is shown, but there are four IHXs immersed in the primary coolant located in the Hot Pool annular region between the cylindrical Inner Vessel and Reactor Vessel. No solid structure separates the primary coolant hot and cold collectors.

The Reactor Vessel is a cylindrical shell with a hemispherical bottom.

The Internal Structure is not axially symmetrical but has a transversal symmetry plane (Fig. 3). The IHXs are of the "bayonet" type, with arrays of straight tubes contained within a vertical shell, anchored to an upper support tube sheet. A single bayonet assembly consists of a pair of concentric tubes to allow the inversion of the secondary coolant flow from downward to upward (Fig. 2). The IHX is enclosed in a 1 cm steel shell, except for the upper 60 cm, which serve as inlet region. The skirt (lower part of the shell) goes 1.4 m deeper than the tubes to enhance the natural circulation driving force.

## 2. NUMERICAL SIMULATION GENERAL CHARACTERISTICS

The computational domain used for the CFX simulation is evidenced by the outline shown in Fig. 3 on the mechanical drawing cross sections.

The simulation involves only about one eighth of the whole downcomer.

This approximation has been made due to the limited amount of computational resources (Silicon Graphics OCTANE: RAM, 1 GB; CPU MIPS R12000, 300 MHz), which is a very critical point for this analysis.

A symmetry boundary condition has been imposed on the cutting sections.

The computational model is presented in details in Fig. 4. The names of the components mentioned below refer to the name of the components in this figure.

The Pb-Bi eutectic enters the domain from the top cross section (yellow) of the vertical rising pipes (red) and leaves the domain at the bottom part through an Internal Support Holes (green). Then it turns back to the core. This except for a small part of the flow, which leaves the domain through the Core Bypass Flow Holes (magenta). The red and blue parts of the grid represent the IHX. At the level of the unwrapped red grid the flow enters the IHX. The blue part of the grid is the one wrapped by the shell, while the grey part is the skirt which guides the flow vertically under the IHX. The cyan part is the secondary coolant (oil) feeding tube. The upper parts of the domain coincide with the Pb-Bi eutectic free surface.

In the computational model the same co-ordinate system as in Fig. 3 has been used. The geometrical characteristics of the model are as follows:

| | |
|---|---|
| Reactor Vessel external radius / thickness: | 3 m / 0.04 m |
| IHX centreline radial co-ordinate: | 2.39 m |
| Inner Vessel Low Part external radius / thickness: | 1.9 m / 0.02 m |
| Inner Vessel Upper Part external radius / thickness: | 1.59 m / 0.02 m |
| Internal Support radius : | 1.56 m |

The axial co-ordinates of the main components of the model are:

| | |
|---|---|
| Free surface level: | -2.3 m |
| Riser pipes top, Oil Tube top: | -2.6 m |
| Oil Tube bottom: | -3.2 m |
| IHX Shell top ($z_{IHXshell}^{top}$): | -3.8 m |
| Connection Plate (Risers Support) | -6.425 ÷ -6.475 m |
| Core Bypass Flow Holes centre | -7.1 m |
| IHX Outlet (bottom) ($z_{IHX}^{bottom}$): | -7.3 m |
| Reactor Vessel Spherical Part centre: | -8.16 m |
| IHX Shell Outlet: | -8.7 m |
| Core Diagrid | -9.48 ÷ -9.63 m |
| Internal Support Holes centre | -9.92, -10.13, -10.34 m |
| Model bottom: | -10.63 m |

The computational domain inlet consists of the outlet of 2 rising pipes. The area of the simulated pipes has been adjusted due to the fact that the riser tubes numbers for each IHX is 6 rising pipes (3 rising pipes for half IHX).

As a consequence the pipes in the computational domain have been transformed into pipes with an ellipsoidal cross section with semiaxis 10 cm x 12.61 cm instead of a circular cross section with radius of 10 cm.

The inlet-imposed velocity is about 0.88 m/s, which gives the desirable mass flow rate. This velocity is a slightly higher than the expected mean velocity in the real rising pipes, but it is consistent with a bubbly flow with a void fraction of about 14% at the rising pipes outlets. The regions with bubbly flow rising pipes have not been included in the computational domain to save computer resources. The model includes only external surfaces of the rising pipes.

The inlet turbulent intensity $\tau_{inlet}$ and the turbulent length $l_{inlet}$ must be specified when applying the RNG κ–ε turbulence model. We assigned $l_{inlet}$ = 0.2 m, equal to the inner diameter of the riser tubes, and $\tau_{inlet}$ = 0.037.

The inlet temperature of the computational domain is $T_{inlet}$ = 400 °C.

The computational domain outlet consists of three annular slots in the Internal Support. They are representative of the three series of holes in the real geometry.

Pb-Bi eutectic lives the domain also through an annular slot in the Inner Vessel, which represents the Core Bypass Flow Holes in the real geometry. An inlet boundary condition with negative velocity has been used at this boundary.

The appropriate velocity value has been chosen to obtain the design value of the core bypass flow rate. The height of each annular slot is set to the minimum value that allows four cells in the axial direction.

The free surface at the top of the coolant is not simulated precisely. In most runs the free surface was represented by a symmetry plane. This is for a number of reasons ranging from the additional computational cost associated with the free surface simulation to the fact that in reality the riser outlet is characterised by a bubbly flow regime. It has anyway to be pointed out that the effect of the presence of the free surface is expected to be localised in the upper part of the domain and should have a minimum influence on the thermal-hydraulic field in the rest of the domain. In fact, some runs using different free surface models (solid wall or moving free surface in the transient run) produced very similar results, at least in the regions far from free surface.

The Reactor Vessel Spherical Part has been simulated using CFX User Fortran routines, introducing a region with a very high resistance in place of a solid wall (Fig. 4).

The IHX has been simulated as a porous media. The porosity is constant and is taken directly from the bundle geometrical structure. The relevant IHX characteristics used by the numerical simulation are given below:

Outer tube external diameter: $d = 25.4$ mm  
Outer tube thickness: $t = 1.2446$ mm  
Inner tube external diameter: $d_1 = 19.05$ mm  
Number of tubes: $n_{tubes} = 1072$  
Number of rods: $n_{rods} = 24$  
Pitch: $s = 32.5$ mm  
Roughness of surfaces: $\Delta = 0.05$ mm  

Hydraulic diameter: $d_h = 20.45$ mm  
Porosity: $\alpha = 0.4666$  
Superficial area: $1.041$ m$^2$  
Height: $4.1$ m  
IHX shell external dimensions:  
675 x 1750 x 4900 mm  

The porous media resistance is modelled in CFX User Fortran routines as additional body forces in the momentum equations and comes from pressure drop correlations applied locally. Different correlations have been used for the axial and cross flow directions in the bundle. Localised pressure losses have been also taken into account. In particular localised pressure losses of the sets of sustaining grids have been simulated. Each set, has a pressure loss coefficient $\varsigma$ of 0.23 based on the actual velocity inside the IHX. We have considered three of these sets. Numerically, the corresponding losses have been distributed on the part of the IHX enclosed in the shell.

The heat exchange in IHX between primary and secondary coolant has been simulated as a source term in the enthalpy equation using CFX User Fortran routines. The Pb-Bi eutectic side heat exchange coefficient are calculated locally, while oil side heat exchange coefficient is assumed as constant. In order to simulate the secondary coolant behaviour IHX has been logically subdivided in two parts. In each part the oil temperature changes linearly with axial co-ordinate (see Fig. 8). The slope of the changes depends on the heat transmitted from Pb-Bi eutectic to oil and has been calculated in each iteration. The Oil inlet temperature is also calculated in each iteration to provide a boundary condition fulfilling the correct power balance of the system.

Apart from the primary-secondary coolant heat transfer, many other heat exchanges have been taken into account across walls in the computational domain. The appropriated features of the code have been used to implement these boundary conditions. In the following a short description of the heat transfer implementation is given.
The location of the walls is shown in Fig. 4.

Conducting walls, which has been actually meshed through their thickness are: IHX Shell, Reactor and Inner Vessels, Connection Plate (or Risers Support), Core Diagrid.
Conjugate heat transfer is calculated on the surfaces of these walls bounded by Pb-Bi eutectic. The X-ADS can release heat through a safety device called RVACS, which is external to the Reactor Vessel and is based on the natural convection of external air. The heat flux, q'', depends on the Reactor Vessel outer wall temperature ($t_w$, expressed in $^oC$) according to the following law:

$$q'' = -158. + 1.57*t_w - 1.52*10^{-2}*t_w^2 \text{ in W/m}^2.$$

This law is used to set the local heat flux on the Reactor Vessel outer wall.

The heat flux through the walls listed below are simulated by setting an external heat exchange coefficient (or thermal resistance $r_{th}$) and an external temperature.

- Risers Support bottom wall, Inner Vessel Lower Part 1 internal wall:
  external temperature 400 $^oC$, thermal resistance $r_{th} = 1.4*10^{-4}$ m$^2$K/W.
- Inner Vessel Lower Part 2 internal wall, Core Diagrid top wall:
  external temperature 300 $^oC$, thermal resistance $r_{th} = 1.8*10^{-2}$ m$^2$K/W.

The following structures have zero thickness in the model: Riser pipes, Oil Tube, Oil Tube top and bottom, Oil Tube Cilindrical Part, Internal Support.

The heat flux through the walls listed below are simulated by setting a fixed external temperature with a given thermal resistance $r_{th}$ corresponding to the wall characteristics (conductivity and thickness) and external heat exchange coefficient.

- Risers internal walls: external temperature 400 $^oC$ with $r_{th} = 1.4*10^{-4}$ m$^2$K/W.
- Lateral and top of IHX feeding tube connection (Oil Tube, Oil Tube Top):
  external temperature 320 $^oC$ with thermal resistance $r_{th} = 8.9*10^{-4}$ m$^2$K/W.
- Oil Tube Cylindrical Part: external temperature 320 $^oC$ with $r_{th} = 2.1*10^{-3}$ m$^2$K/W.

All other walls are treated as adiabatic.

3. MAIN SIMULATION PARAMETERS

The main parameters of the primary circuit are reported below.
The values of the parameters are reported for one IHX, for half of the IHX effectively simulated, and for the whole vessel.

Primary coolant: Lead-Bismuth Eutectic
Primary coolant mass flow rate: $M_1 = 4 \times 1450 = 8 \times 725 = 5800$ kg/s
Secondary coolant: Diphyl THT Oil
Secondary coolant mass flow rate: $M_2 = 4 \times 206.7 = 8 \times 103.35 = 827$ kg/s
Nominal power: $P_w = 4 \times 20 = 8 \times 10 = 80$ MW

The following coolant properties have been used for the analysis.

Reference temperature for Lead-Bismuth eutectic fluid and steel: $t_{ref}$ = 350 °C

| **Properties** | **Lead-Bismuth eutectic** | **Diphyl THT** |
|---|---|---|
| Density: | $\rho$ = 10270 kg/m³ | $\rho_{oil}$ = 822 kg/m³ |
| Thermal exp. coefficient: | $\beta$ = 1.216 · 10⁻⁴ 1/K | |
| Laminar kinematic viscosity: | $\nu$ = 17.38 · 10⁻⁸ m²/s | |
| Laminar dynamic viscosity: | $\mu$ = 1.784 · 10⁻³ kg/m/s | $\mu_{oil}$ = 4.6 10⁻⁴ kg/m/s |
| Thermal conductivity: | k = 13.2 W/m/K | $k_{oil}$ = 0.102 W/m/K |
| Specific heat: | $C_p$ = 146 J/kg/K | $C_{p\,oil}$ = 2510 J/kg/K |

Steel properties:
Density: $\rho_{steel}$ = 7824 kg/m³
Thermal conductivity: $k_{steel}$ = 18.67 W/m/K
Specific heat: $C_{p\,steel}$ = 539 J/kg/K

## 4. RESULTS OF THE CFD ANALYSIS

Several runs using different meshes have been performed. The following discussion is based on one "reference" run of our analysis, although main results of other simulations are briefly cited when strictly needed.

Fig. 3 presents the domain of the calculation with reference to the reactor block assembly. The grid used in the final calculation is shown in fig. 4 and 5 where the main components are detailed. This grid is characterised by about 800 thousands nodes.

The grid nodes have been distributed in order to obtain a sufficient level of detail in the zone of interest for the calculation. Values of the $y^+_{nw}$ (non-dimensionalised distance from wall of the nearest-to-wall cell center) have also been checked with the aim to verify that they fall into the range of values suitable for the simulation using the wall function concept.

The zone were thermal stratification takes place is located at the exit of the IHX skirt. In this region the flow coming from IHX mixes with the flow bypassing it, with a consequent generation of a thermal stratification phenomena.

Figures 6 and 7 presents the result of the calculation in terms of velocity field and temperature distribution respectively. Such results are related to a steady state mode of the numerical calculation. It has to be noted that all calculations we performed using different grid sizes and time steps in both steady state (false time steps) and full transient mode of the numerical calculation confirmed that the main thermal stratification gradient is located at the outlet of the IHX skirt, when the amount of the IHX by-pass flow is higher then 4% of the total flow (note that the expected design conditions are close to 7-8%).

In particular the gradient was also found stable restarting the calculation in full transient mode from steady state conditions although some oscillations in terms of maximum gradient elevation were found.

Such oscillations have not been analysed in detail due to the inherent limitation of the model used for the simulation. They are justified by the rather unstable nature of the vortices developed in the upper and lower part of the downcomer, which interacts at the thermal stratification level. Fig. 6 shows clearly the complicated flow pattern developed in both the upper and lower part of the downcomer.

Fig. 10 shows the behaviour of the temperature as a function of time for the downcomer average temperature at the IHX exit and two local reactor vessel temperatures. This behaviour suggests oscillations of the layer with high thermal gradient. It has to be noted that these oscillations of the thermal gradient location may give rise to additional thermal loads.

Fig. 7 presents the temperature distribution on the reactor and inner vessel and puts in evidence the rather small extension of the region with high thermal gradient as shown in detail on Fig. 8. Such extension can be easily evaluated.

Taking into account that the maximum temperature difference at the location of the thermal gradient is about 60 K (Fig. 8), and the maximum temperature gradient is about 600 K / m (Fig. 9), we obtain roughly an axial extension of the region with high thermal gradient of 10 cm.

Some final considerations may be added to the presented results taking into account the sensitivity analysis already performed on the grid sizes and on the amount of the by-pass flow.

We observed an important dependence of the thermal stratification from axial and radial grid sizes and this fact was the main reason for the use of the as fine as possible grid presented. We also investigated the behaviour of the thermal stratification as a function of IHX by-pass flow. The main results of this investigation is that the thermal gradient is characterised by a maximum when the total by-pass flow is about 7% of the total flow (riser flow) of the X-ADS facility.

As a consequence we decided to present here as a reference case of our calculation what we considered the worst case from the point of view of the thermal loads on structures.

## 5. CONCLUSIONS

All performed calculations identified the thermal stratification gradient location on the same point of the calculation domain, i.e. at the exit of the skirt of the IHX, if the amount of the IHX by-pass flow is higher then 4% of the total flow (note that the expected design conditions are close to 7-8%).

The amount of thermal stratification was found to be strongly dependent from the IHX by-pass flow. The final solution presented has been checked in terms of stability (transient calculations predicts the same phenomena of steady state calculations) and accuracy ($y^+_{nw}$ parameter was maintained into its acceptable range).

The maximum axial gradient of temperature at wall/fluid interface at reactor and inner vessels achieved in the calculation was of about 600 K / m corresponding to a temperature difference of about 60 K located in a region of 10 cm thickness.

The results should however be interpreted taking into account the following considerations:

- The calculation was performed on a relatively small slice of the X-ADS downcomer.
- The stability of the thermal stratification gradient has been analysed only qualitatively by transient calculations showing that the oscillation of the thermal stratification could increase the thermal loads respect to that calculated by a steady state analysis.

Having in mind the above limitations of the present calculations we can say that:

- The existence of a thermal stratification gradient has been confirmed by the calculations, its location and value determined.
- The steady state thermal loads as calculated in this analysis have been found acceptable for the reactor and inner vessels.
- In the case that more refined thermal-hydraulic analysis give rise to unacceptable loads, the introduction of adequate design modifications (for example thermal shields) may be chosen as a final solution.

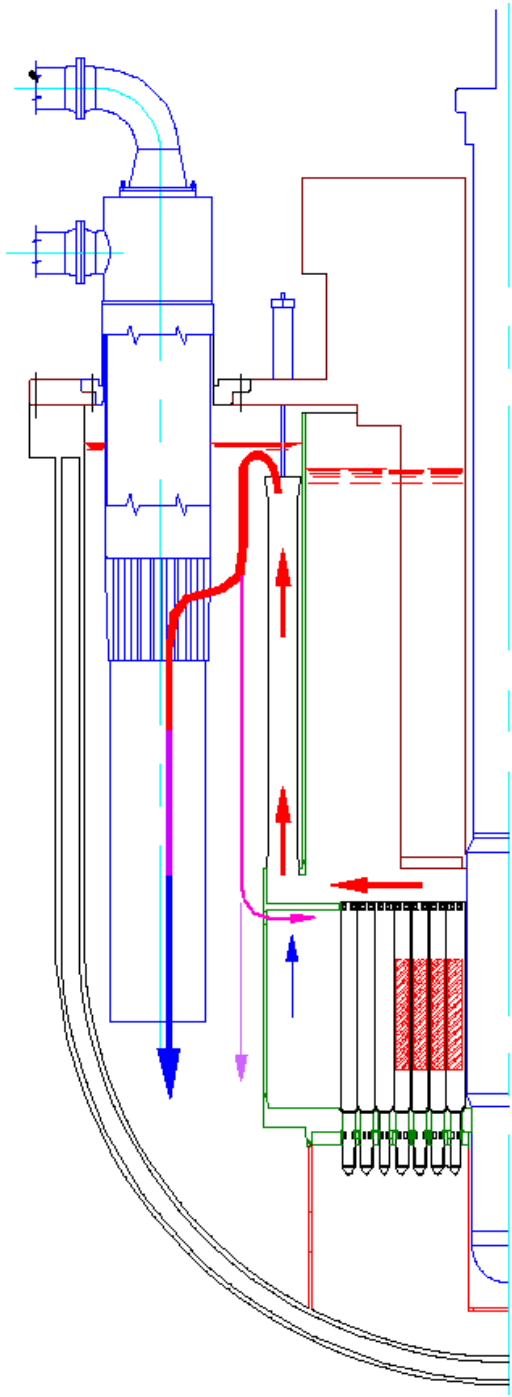

Fig. 1 - Primary coolant flow paths

Fig. 2 - Intermediate Heat Exchanger

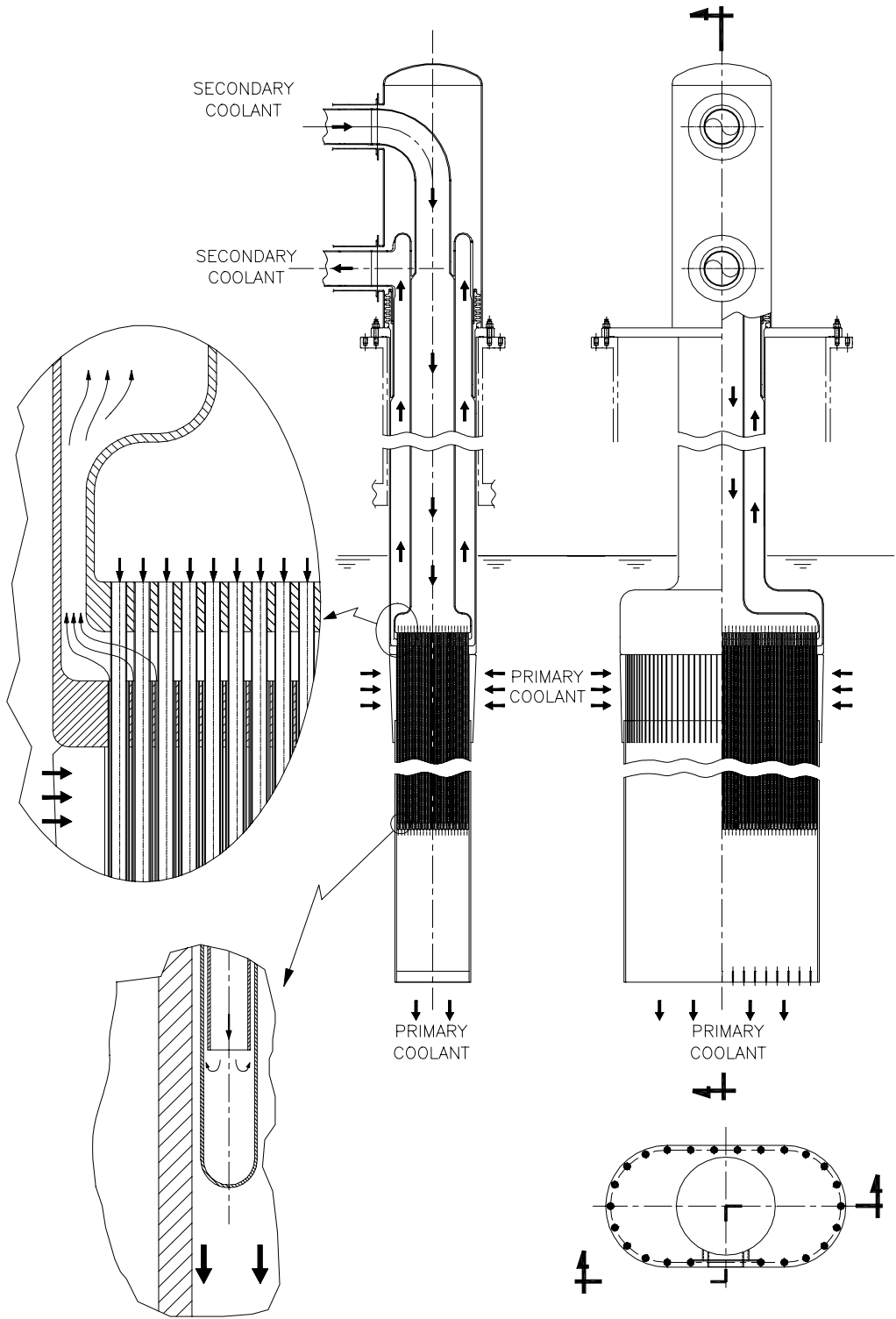

Fig. 3 - CFX Computational domain

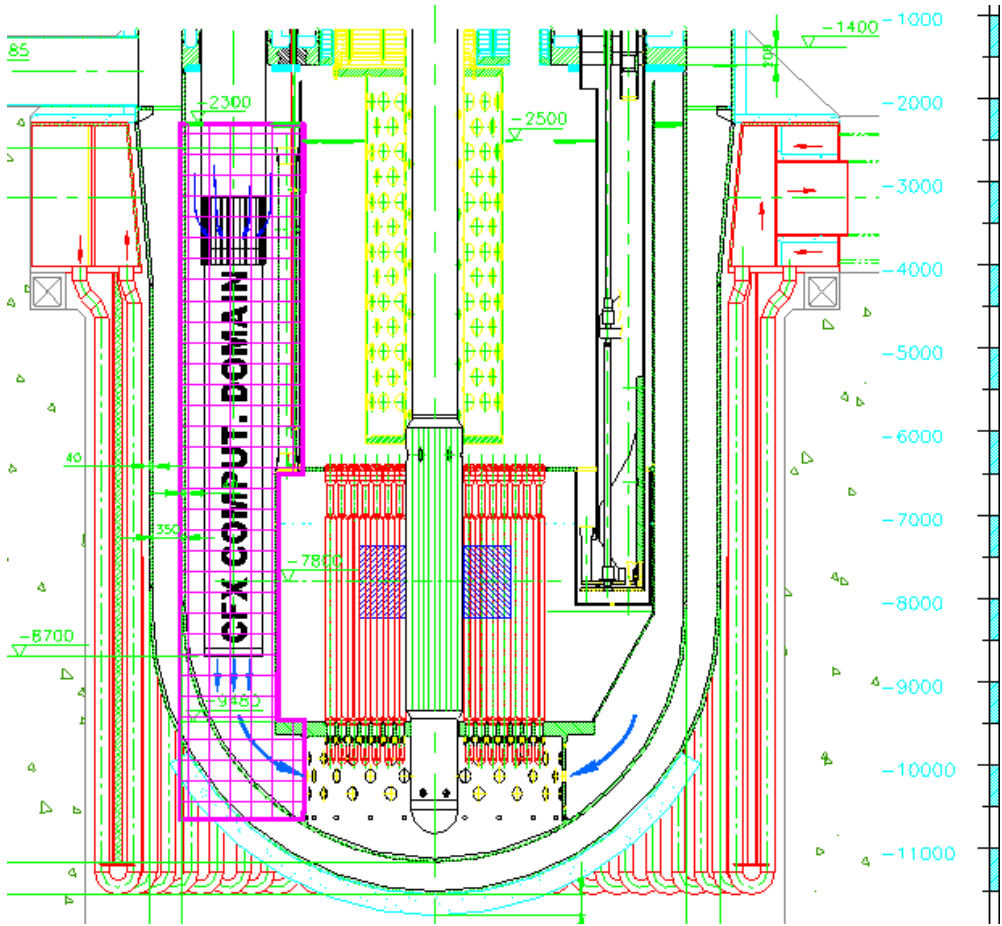
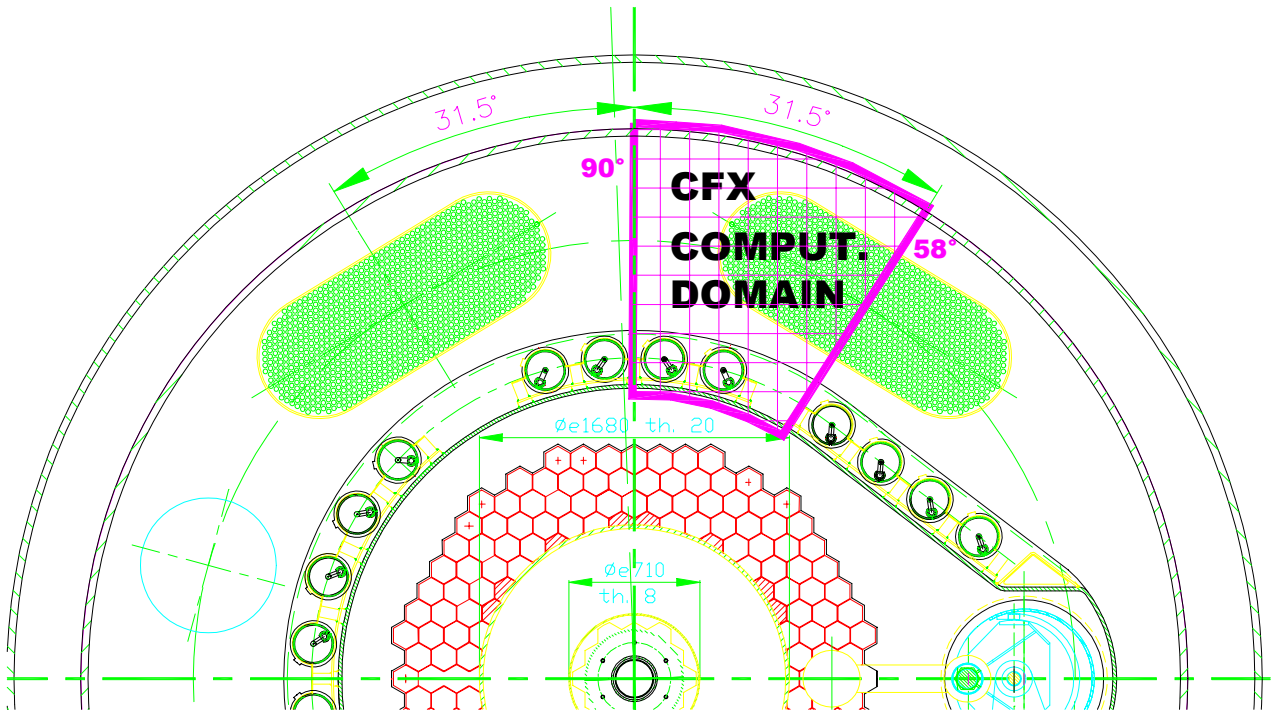

Fig. 4 - CFX DOWNCOMER MODEL

Model isometric view

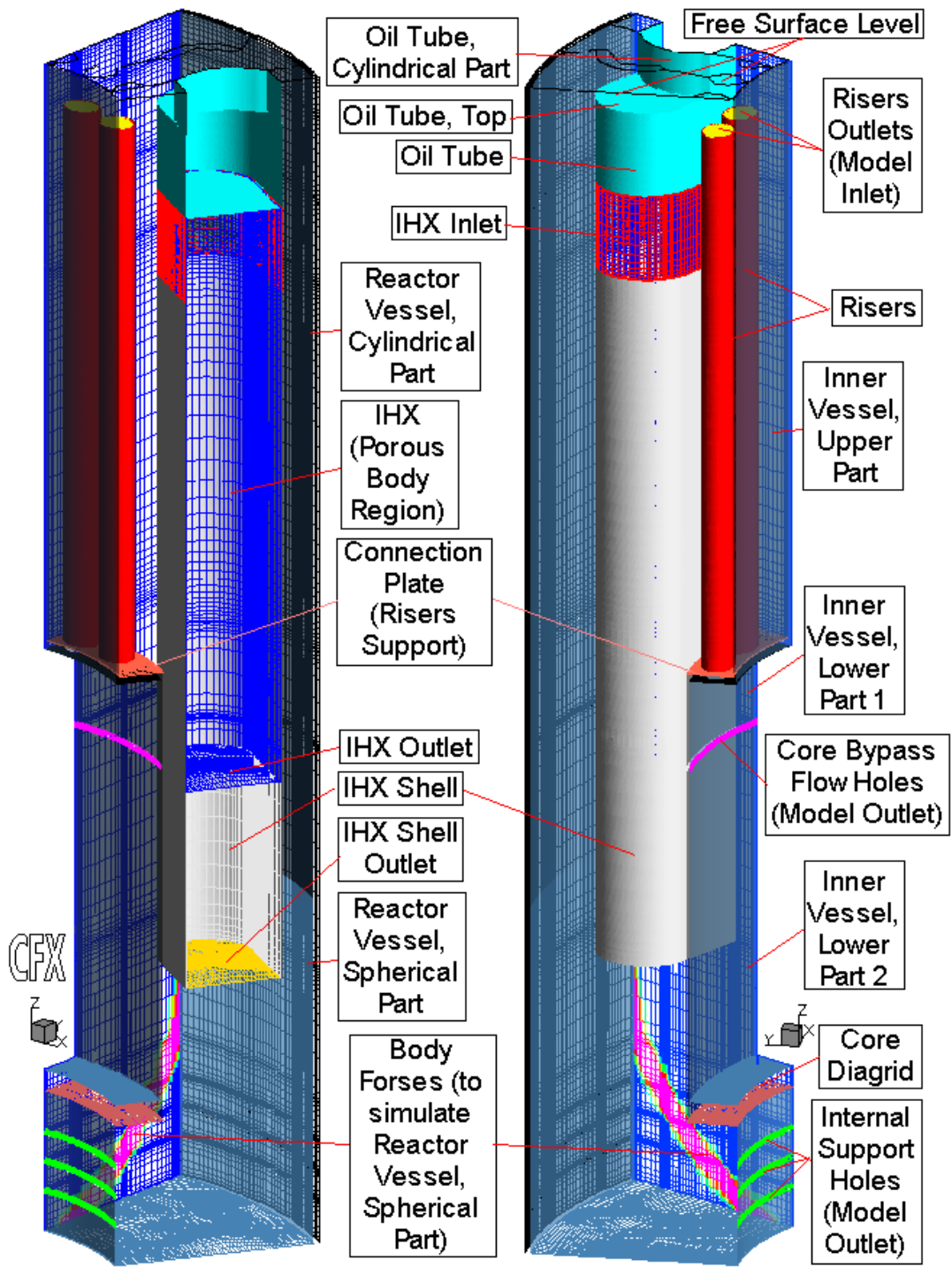

Fig. 5 - CFX DOWNCOMER MODEL

Mesh top view

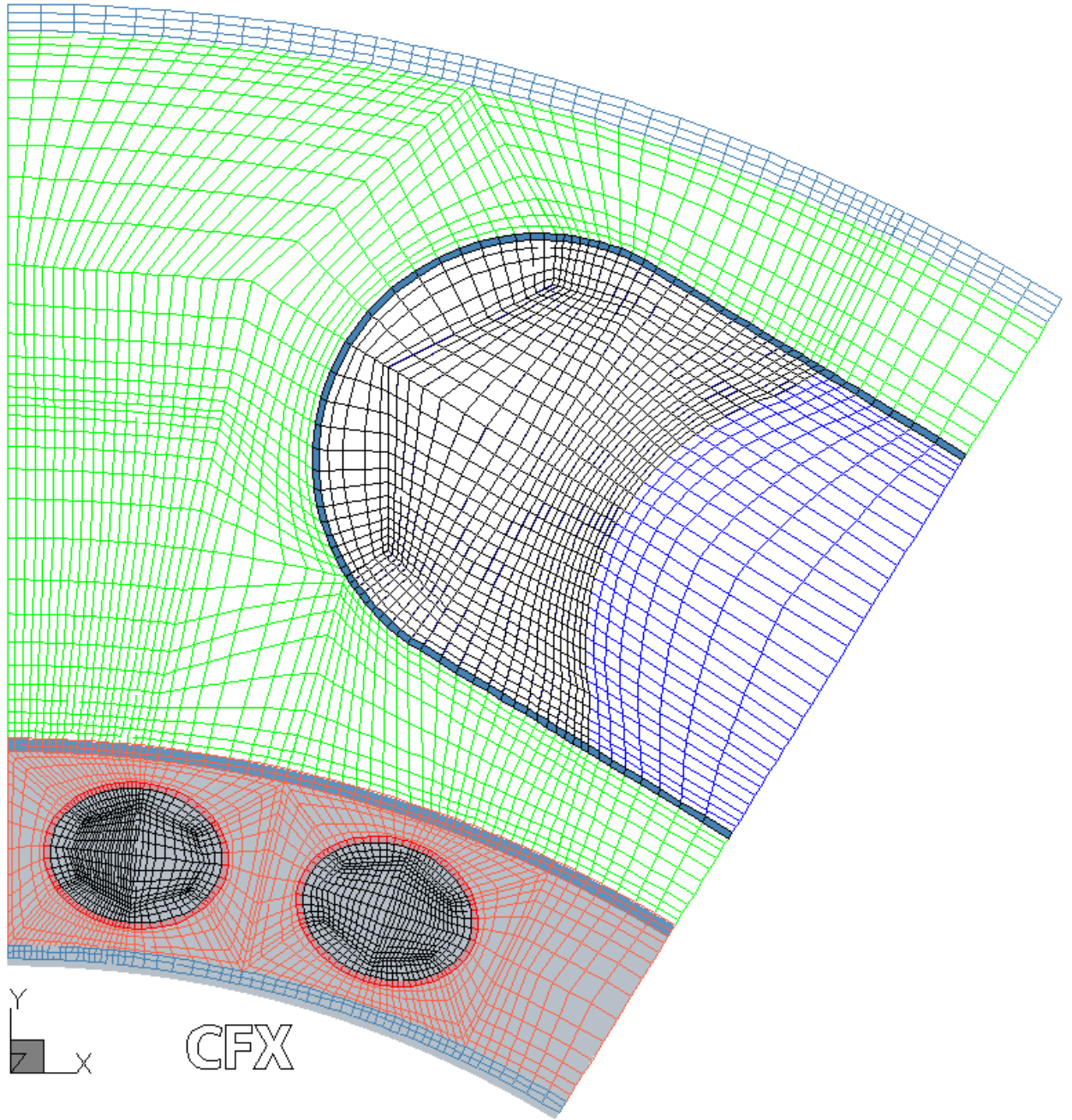

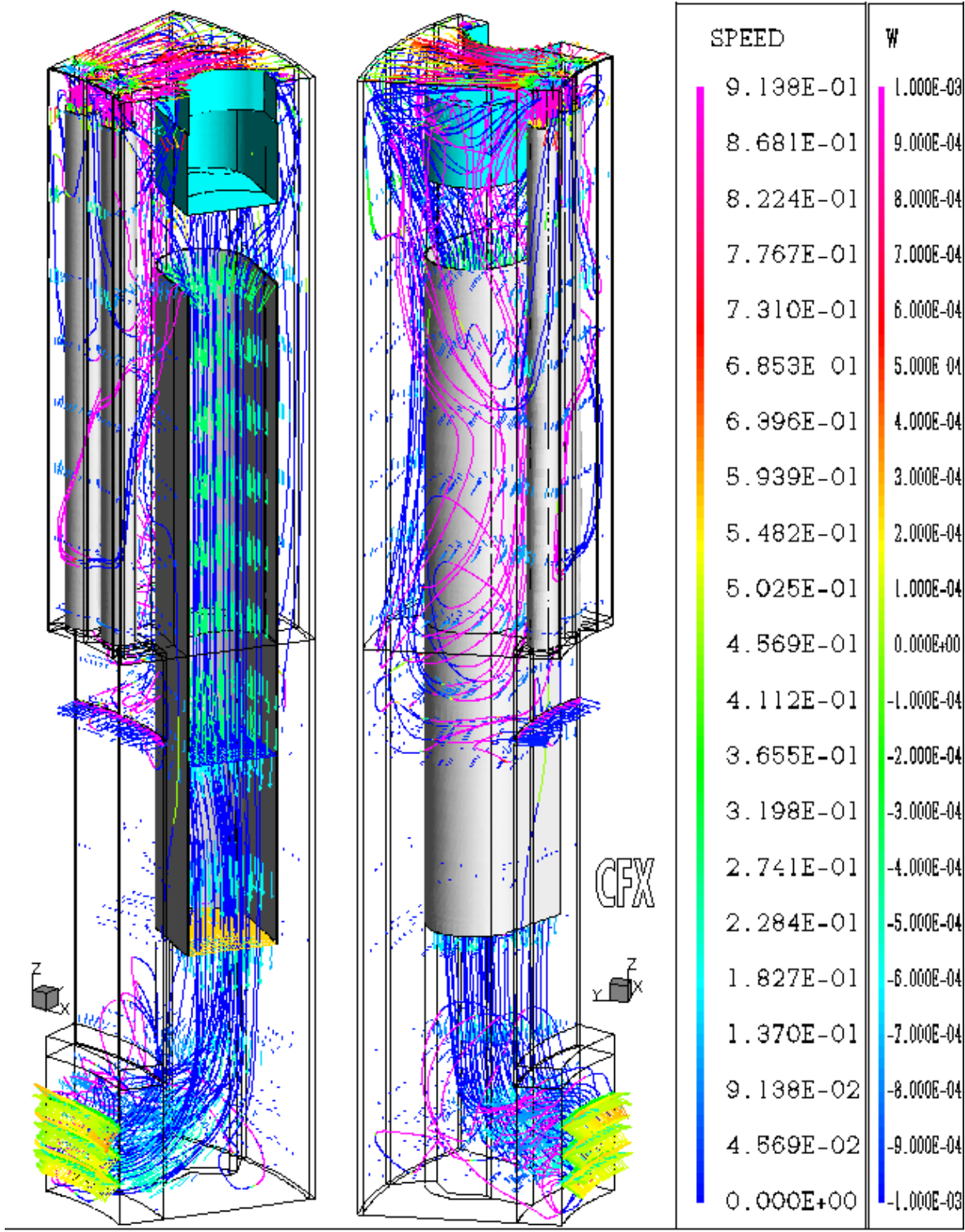

Fig. 6 - VELOCITY FIELD

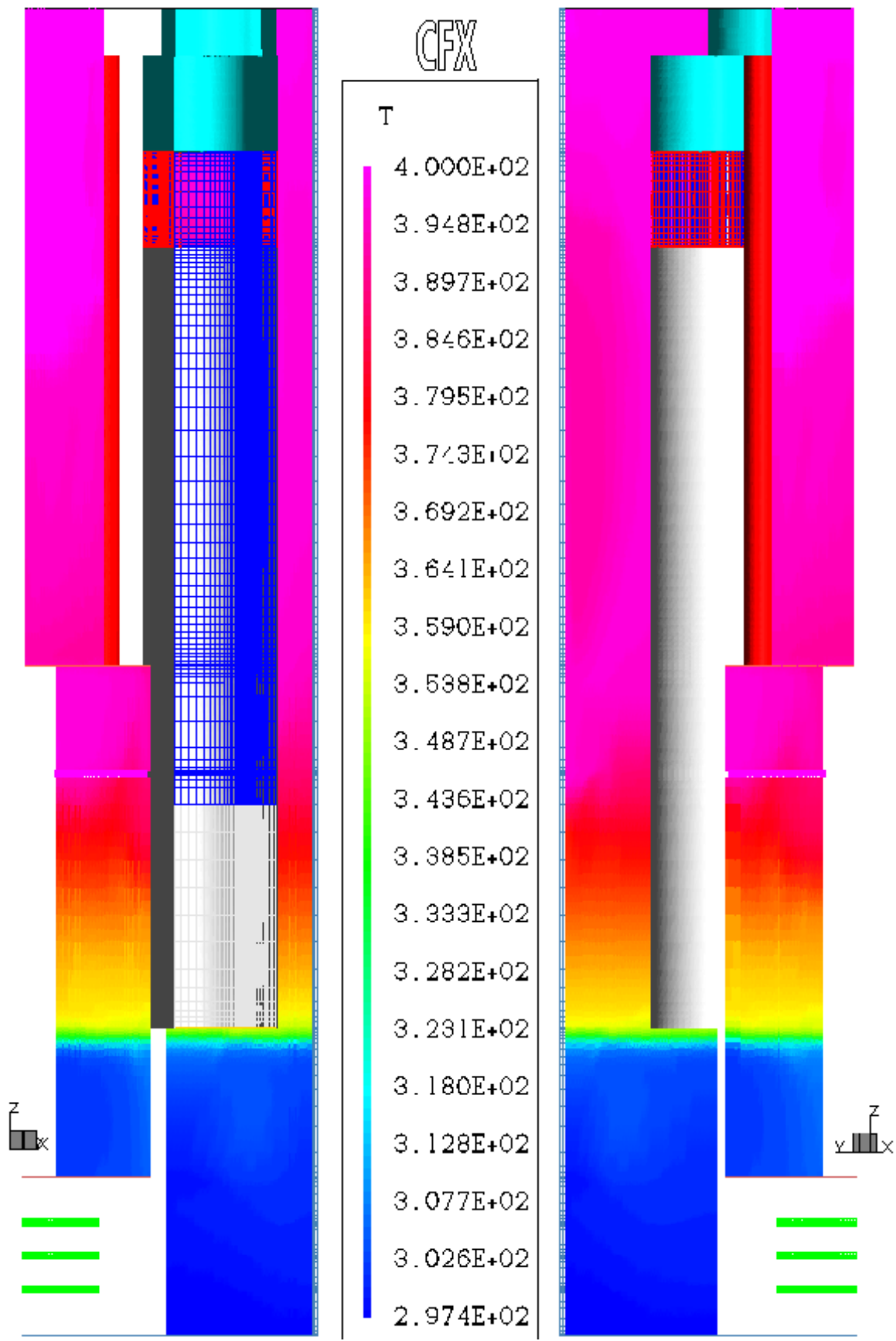

Fig. 7 - VESSELS TEMPERATURE DISTRIBUTION

Fig. 8 - AXIAL TEMPERATURE DISTRIBUTION

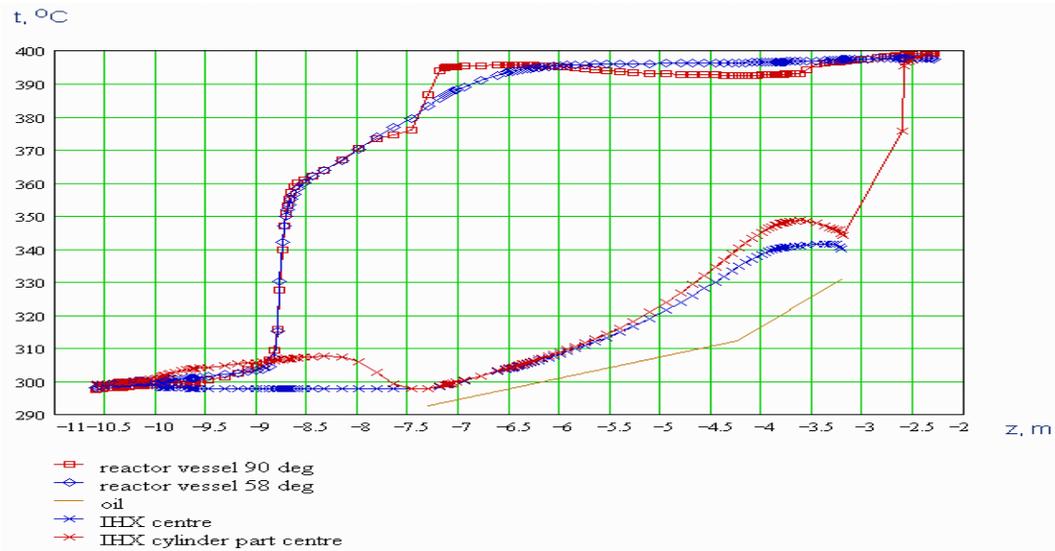

Fig. 9 - AXIAL THERMAL GRADIENT DISTRIBUTION

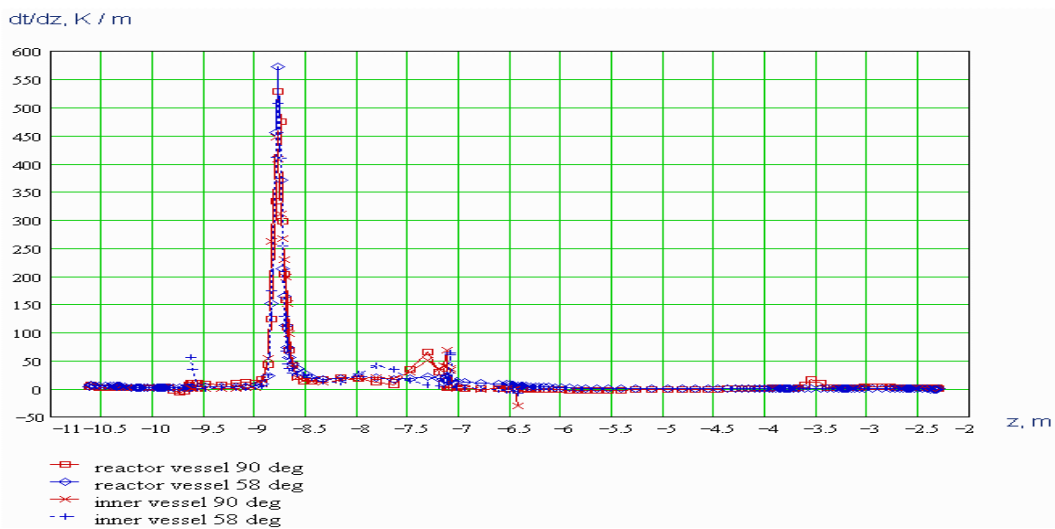

Fig. 10 - TRANSIENT TEMPERATURE OSCILLATIONS

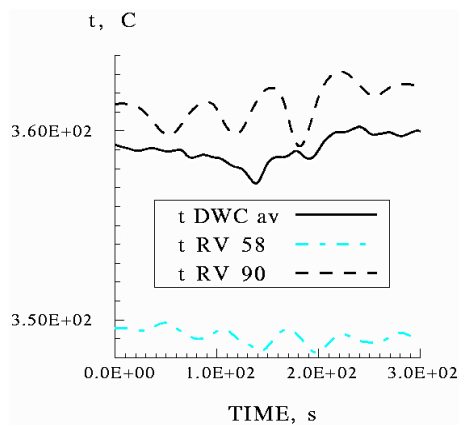